\documentclass[12pt,preprint]{aastex}

\usepackage[]{natbib}

\newcommand{\rxte}{{\it RXTE}}
\newcommand{\xte}{{\it RXTE}}
\newcommand\cxo {{\it Chandra}}
\newcommand\xmm {{\it XMM-Newton}}
\newcommand{\tfe}{1E~1048.1--5937} 
\newcommand{\tfn}{1E~2259+586}

\newcommand{\ett}{XTE~J1810--197}
\newcommand{\tnin}{\mathrm{T}_{90}}

\def\lapp{\ifmmode\stackrel{<}{_{\sim}}\else$\stackrel{<}{_{\sim}}$\fi}
\def\gapp{\ifmmode\stackrel{>}{_{\sim}}\else$\stackrel{>}{_{\sim}}$\fi}

\begin{document}

\title{A Burst and Simultaneous Short-Term Pulsed Flux Enhancement from the
Magnetar Candidate 1E~1048.1--5937} 

\author{Fotis P. Gavriil, Victoria M. Kaspi}
\affil{Department of Physics, Rutherford Physics Building,
McGill University, 3600 University Street, Montreal, Quebec,
H3A 2T8, Canada}
\and
\author{Peter M. Woods\altaffilmark{1}}
\affil{Space Science Research Center, National Space Science
and Technology Center, Huntsville, AL 35805, USA}
\altaffiltext{1}{Universities Space Research Association}

\begin{abstract}
We report on the 2004 June 29 X-ray burst detected from the direction
of the Anomalous X-ray Pulsar (AXP) \tfe\ using the \textit{Rossi
X-ray Timing Explorer} (\xte).  We find a simultaneous increase of
$\sim3.5$ times the quiescent value in the 2--10~keV pulsed flux of
\tfe\ during the tail of the burst which identifies the AXP as the
burst's origin.  The burst was overall very similar to the two others
reported from the direction of this source in 2001. The unambiguous
identification of \tfe\ as the burster here confirms it was the origin
of the 2001 bursts as well. The epoch of the burst peak was very close
to the arrival time of \tfe's pulse peak.  The burst exhibited
significant spectral evolution with the trend going from hard to
soft. Although the average spectrum of the burst was comparable in
hardness ($\Gamma\sim 1.6$) to those of the 2001 bursts, the peak of
this burst was much harder ($\Gamma \sim 0.3$).  During the 11 days
following the burst, the AXP was observed further with \xte, \xmm\ and
\cxo.  Pre- and post-burst observations revealed no change in the
total flux or spectrum of the quiescent emission. Comparing all three
bursts detected thus far from this source we find that this event was
the most fluent ($>3.3\times 10^{-8}$~erg~cm$^{-2}$ in the 2--20~keV
band), had the highest peak flux ($59\pm9\times
10^{-10}$~erg~s$^{-1}$~cm$^{-2}$ in the 2--20~keV band), and the
longest duration ($>699$~s). The long duration of the burst
differentiates it from Soft Gamma Repeater (SGR) bursts which have
typical durations of $\sim$0.1~s.  Bursts that occur preferentially at
pulse maximum, have fast-rises and long X-tails containing the
majority of the total burst energy have been seen uniquely from
AXPs. The marked differences between AXP and SGRs bursts may provide
new clues to help understand the physical differences between these
objects.

\end{abstract}


\section{Introduction}
\label{sec:intro}

The nature of the source class colloquially known as Anomalous X-ray
Pulsars (AXPs) has been well explained by the so called magnetar model
\citep{td96a}. In this model, AXPs, which appear to be isolated, have
pulse periods in the narrow range of 5--12~s, and X-ray luminosities
which greatly exceed their spin-down luminosities, are young,
isolated, ultra-magnetized neutron stars which are powered by their
decaying magnetic fields. The magnetar model was first introduced to
explain the enigmatic properties of the Soft Gamma Repeaters
(SGRs). The SGRs were first identified by their emission of long-lived
super-Eddington flares in soft $\gamma$-rays and by their much more
frequent and less energetic short bursts in hard X-rays
\citep{mgi+79,hcm+99}.  Since their discovery, the quiescent emission
of SGRs has been shown to possess numerous similarities with the
persistent emission of AXPs. Specifically, the SGRs were shown to have
similar pulse periods and spin down rates \citep{kds+98, ksh+99}. For
a recent review of AXPs and SGRs see \citet{wt04}.  

The distinction between AXPs and SGRs was blurred even further when
the \textit{Rossi X-ray Timing Explorer} (\rxte) discovered two
magnetar-like X-rays bursts from the direction of AXP \tfe\
\citep{gkw02}.  The temporal, energetic and spectral properties of
these bursts were very similar to those of SGRs.  Due to the large
field of view of \xte, the AXP could not be unambiguously be confirmed
as the source of the bursts, however the burst properties argued
strongly that the AXP was indeed the origin. The issue of whether AXPs
emit bursts was settled unambiguously when another AXP, \tfn, underwent a major
SGR-like outburst involving 80 bursts accompanied by severe changes to
every aspect of the pulsed emission \citep{kgw+03, wkt+04}.  The
distribution of durations and fluences of these bursts were very
similar to the ones seen in SGRs \citep{gkw04}.  Despite the many
similarities between SGR and AXP bursts, there were some
differences. For instance, the first burst detected from the direction
of \tfe\ exhibited an unusual spectral feature near $\sim14$~keV
during the first $\sim 1$~s of the burst.  Furthermore, in AXPs the
more luminous bursts had harder spectra, the opposite of what is seen
in SGRs.

Besides being able to account for the quiescent emission of SGRs and
AXPs, the magnetar model provides a clear explanation for the bursts
\citep{td95}.  The highly twisted interior magnetic fields of a
magnetar diffuses out and unwinds, causing enormous stresses on the
neutron star's crust. When the crust yields to these internal
stresses, the well anchored foot-points of the external field are
displaced, sending electromagnetic disturbances, i.e. Alfv\'en waves
into the magnetosphere. The Alfv\'en waves provide the energy and
momentum required for photons in the magnetosphere to pair-produce,
yielding an electron-positron and photon fireball.  This fireball is
contained by closed field lines, and is gradually radiated away.  The
magnetic field strengths required to contain the energy released in
SGR bursts are consistent with the dipolar field strengths inferred
from their spin-down \citep{kds+98, ksh+99}.

We report here, using data from our continuing \xte\ monitoring
program, the discovery of another X-ray burst from AXP \tfe.  We show
that the pulsed emission increased in the tail of the burst,
indicating that the AXP was unambiguously the source of the burst. The
identification of \tfe\ as the burst source for this event and the
similarities between this burst and the previous two lends further
support to the AXP having also been the emitter of the two bursts
reported by \citet{gkw02}.

\section{Results}
\label{sec:results}

\subsection{\xte\ Observations}

The results presented here were obtained using the Proportional
Counter Array \citep[PCA;][]{jsg+96} on board \xte. The PCA consists
of an array of five collimated xenon/methane multi-anode proportional
counter units (PCUs) operating in the 2--60~keV range, with a total
effective area of approximately $\rm{6500~cm^2}$ and a field of view
of $\rm{\sim 1^o}$~FWHM.  We use \xte\ to monitor all five known AXPs
on a regular basis as part of a long-term monitoring campaign
\citep[see][and references therein]{gk02}.  On 2004 June 29, during
one of our regular monitoring observations (\xte\ observation
identification 90076-02-09-02) that commenced at UT 06:29:28, the AXP
\tfe\ exhibited an SGR-like burst.

\subsubsection{Burst Search}

The burst was identified using the burst searching algorithm
introduced in \citet{gkw02} and described in detail in
\citet{gkw04}. To summarize briefly, time series were created
separately for each PCU using all xenon layers. Light curves with time
bin widths of 1/32~s were created.  The \texttt{FTOOL}s
\texttt{xtefilt} and \texttt{maketime} were used to determine the
intervals over which each PCU was off.  We further restricted the data
set by including only events in the energy range 2--20~keV.  Time bins
with significant excursions from a running mean were flagged and
subject to further investigation.  The observation had total on-source
integration time of 2.0~ks.  There were exactly three PCUs operational
at all times and the burst was equally significant in all three
PCUs. Data were collected in the \texttt{GoodXenonwithPropane} mode,
which records the arrival time (with 1-$\mu$s resolution) and energy
(with 256-channel resolution) of every unrejected xenon event as well
as all the propane layer events.  Photon arrival times were adjusted
to the solar system barycenter using a source position of (J2000)
$\mathrm{RA} = 10^{\rm h}\ 50^{\rm m}\ 07\fs 12$, $\mathrm{DEC} =
-59\degr 53\arcmin 21\farcs 37$ \citep{ics+02} and the JPL DE200 planetary
ephemeris.  Note that following the burst we initiated a 20.2~ks long
\xte/PCA Target of opportunity (ToO) observation on 2004 July 8.
Similarly, on 2004 July 10, we initiated 3 more \xte/PCA ToO
observations which had integration times of 3.6~ks, 15.0~ks and 20.2~ks
respectively. No more bursts or other unusual behavior were seen in
the ToO observations or in any of the monitoring observations since
the burst.

\subsubsection{Burst Temporal Properties}
\label{sec:temporal properties}

\begin{figure}
\plotone{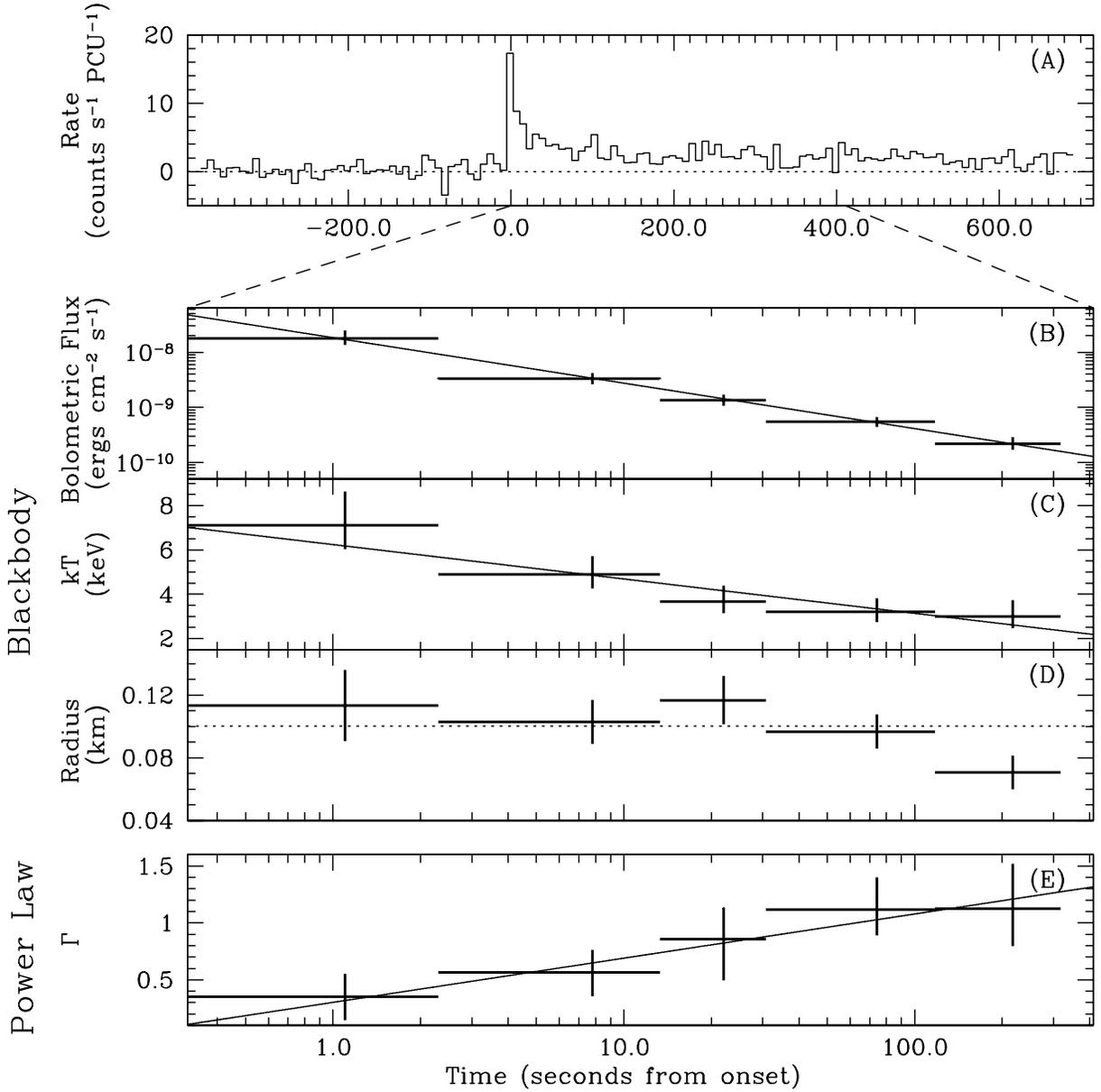} 
\figcaption{A: Background subtracted burst time
series in the 2--20~keV band binned with 8-s time resolution. B: Bolometric
flux ($F$) time series. The line represents the best-fit power law,
$F=1.84\times 10^{-8}(t/\mathrm{1\ s})^{-0.82}$~erg~cm$^{-2}$~s$^{-1}$. C:
Blackbody temperature ($kT$) time series. The line represents the best-fit
logarithmic function, $kT = 6.24 - 1.55\log(t/\mathrm{1\ s})~\mathrm{keV}$.
D: Blackbody radius versus time. The dotted line represents the average
emission radius, $R=0.10$~km assuming a distance of 5~kpc.  E: Power-law
index ($\Gamma$) time series. The line represents the best-fit logarithmic
function, $\Gamma = 0.30 + 0.39\log(t/\mathrm{1\ s})$. \label{fig:profile}}
\end{figure}

We analyzed the temporal properties of the burst in order to compare
them to those of other bursts from AXPs and SGRs.  The analysis
methods are explained in greater detail elsewhere \citep[e.g.,
see][]{gkw02,gkw04,wkg+05} The burst profile is shown in
Figure~\ref{fig:profile} and its measured properties are summarized in
Table~\ref{ta:burst}. The burst peak time was initially defined, using
a time series binned with 1/32~s resolution, as the midpoint of the
bin with the highest count rate. We redefined this value, using the
event timestamps within this time bin, as the midpoint of the two
events having the shortest separation.

Using the burst peak time, we determined the occurrence of the burst
in pulse phase.  We split our observation into four segments and
phase-connected these intervals using the burst peak time as our
reference epoch. We then folded our data using the resulting ephemeris
and cross-correlated our folded profile with a high signal-to-noise
template whose peak was centered on phase $\phi=0$, where $\phi$
is measured in cycles. \citep[For a review of our timing techniques,
see][]{kcs99,klc00,kgc+01,gk02}.  We find that the burst occurred near
the peak of \tfe's pulse profile, at $\phi = -0.078\pm0.016$.  

The burst rise time was determined by a maximum likelihood fit to the
unbinned data using a piecewise function having a linear rise and
exponential decay. The burst rise time, $t_r$, was defined as the time
from the peak to when the linear component reached the background
(\S~\ref{sec:flux and fluence} discusses how the background was
estimated).  The burst duration, $\tnin$, is the interval between when
5\% and 95\% of the total 2--20~keV burst fluence was received.  As we
will note in \S~\ref{sec:flux and fluence}, the burst did not fade away
before the end of our observation. Thus we could only place an upper
limit of $>699$~s on the burst duration, which is the time from the
burst's peak to the end of our observation. This very long tail can be
seen in the burst profile in Figure~\ref{fig:profile}, which shows a
significant excess from the burst's peak to the end of our
observation.

\subsubsection{Burst Spectral Evolution}

Significant spectral evolution has been noted for the first burst
discovered from this source \citep{gkw02} as well as for bursts from
AXP \ett\ \citep{wkg+05} and bursts from SGRs \citep{isw+01,
lwg+03}. Motivated by these observations we extracted spectra at
different intervals within the burst's duration. We increased the
integration time of the spectra as we went further away from the burst
to maintain adequate signal-to-noise. A background spectrum was
extracted from a 1000-s long interval which ended 10~s before the
burst.  From each of the burst intervals and the background interval
we subtracted the instrumental background as estimated from the
tool \texttt{pcabackest}. Each burst interval spectrum was grouped so
that there were never fewer than 20 counts per spectral bin after
background subtraction. The regrouped spectra along with their
background estimators were used as input to the X-ray spectral fitting
software package \texttt{XSPEC}\footnote{http://xspec.gsfc.nasa.gov}.
Response matrices were created using the \texttt{FTOOL}s
\texttt{xtefilt} and \texttt{pcarsp}. All channels below 2~keV and
above 30~keV were ignored leaving 10--24 spectral bins for fitting.
We fit the burst spectra to a photoelectrically absorbed blackbody
model which adequately characterized the data.  In all fits, the column
density was held fixed at the average value of our \cxo\ and {\it
XMM-Newton} observations (see \S~\ref{sec:pep} and
Table~\ref{ta:spectra}).  The burst's bolometric flux, blackbody
temperature and radius evolution are shown in
Figure~\ref{fig:profile}.  The bolometric flux decayed as a power law
in time, $F=F_1(t/\mathrm{1\ s})^{\beta}$, where $F_1 = 1.84 \pm 0.36
\times 10^{-8}$~erg~cm$^{-2}$~s$^{-1}$ and $\beta = -0.82 \pm
0.05$. The blackbody temperature decayed as $kT =kT_1
-\alpha\log(t/\mathrm{1\ s})$, where $kT_1 = 6.24 \pm 0.71$~keV and
$\alpha = 1.55\pm0.41$~keV.  The blackbody emission radius remained
relatively flat with an average value of $R=0.100\pm0.01$~km. The
blackbody radius was calculated assuming a distance of 5~kpc to the
source.  We repeated the above procedure using a power-law model,
which also adequately characterized the data.  Our power-law spectral
index time series is shown in Figure~\ref{fig:profile}, where we see
that the initial spike of the burst is very hard, with the burst
gradually softening as the flux decays.  Fitting a logarithmic
function to the power-law spectral index time series we find $\Gamma =
\Gamma_1 + \alpha\log(t/\mathrm{1\ s})$, where $\Gamma_1 = 0.30 \pm
0.18 $ and $\alpha=0.39\pm0.13$.

\begin{figure}
\plottwo{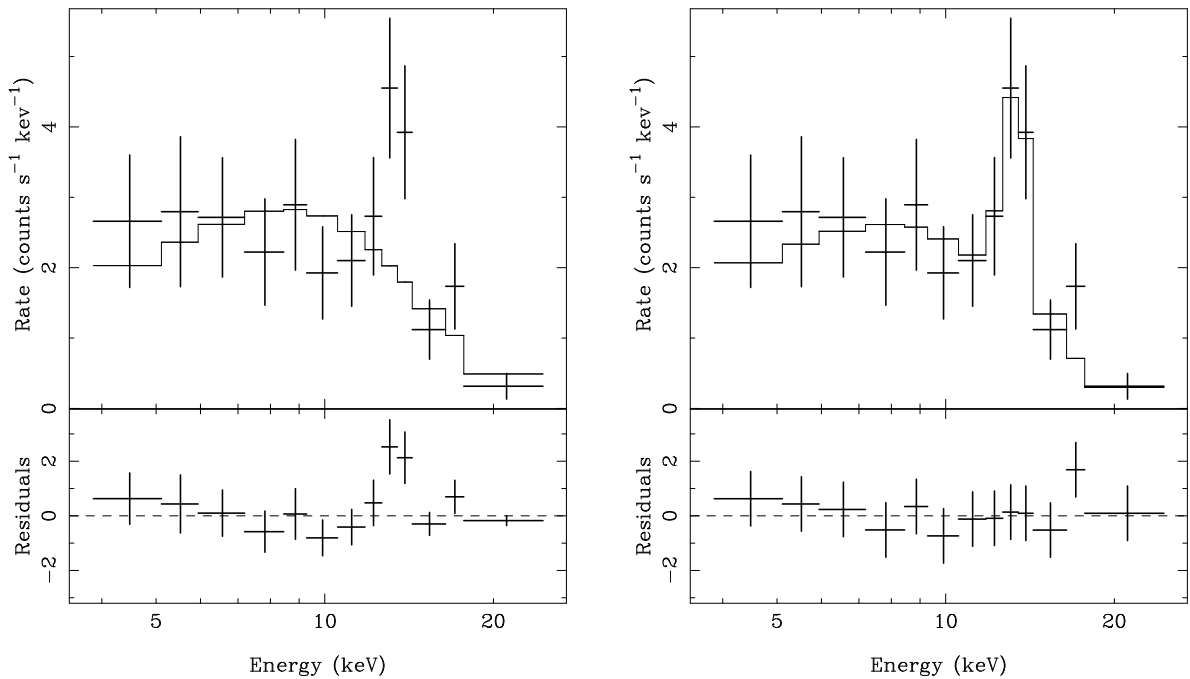}{f3b.eps} \figcaption{Left: An
8-s long spectrum starting 1~s after the peak of the burst fit with a
simple blackbody model. The fit had $\chi_{\mathrm{dof}}=1.6$ for 11
degrees of freedom. There is a possibility of a spectral feature at
$\sim13$~keV. Right: The same spectrum as on the left, but fit with a
blackbody plus a Gaussian emission line. The fit had
$\chi_{\mathrm{dof}}=0.6$ for 8 degrees of freedom.
\label{fig:spectra}}
\end{figure}

Possible spectral features have been reported in bursts from AXPs
\tfe\ \citep{gkw02} and \ett\ \citep{wkg+05} and from bursts from two
SGRs \citep{si00,iss+02,isp03}. We searched for features by extracting
spectra of different integration times as was done for the spectral
evolution analysis.  The spectra were background-subtracted and
grouped in the exact same fashion as in the spectral evolution
analysis.  Energies below 2~keV and above 30~keV were ignored, leaving
13 spectral channels for fitting.  Regrouped spectra, background
estimators and response matrices were fed into \texttt{XSPEC}.
Spectra were fit with a photoelectrically absorbed blackbody model,
holding only $N_H$ fixed at the same value used in the spectral
evolution analysis.  The first 8~s of the burst spectrum was poorly
fit by a continuum model, because of significant residuals centered
near 13~keV.  The apparent line feature was most significant if the
first second of the burst was excluded. A simple blackbody fit had
reduced $\chi^2_{\mathrm{dof}} = 1.61$ for 11 degrees of freedom (see
Figure~\ref{fig:spectra}). The probability of obtaining such a value
of $\chi^2$ or larger under the hypothesis that the model is correct
is very low, $P(\chi^2 \ge 17.75 ) = 0.088$. The fit was greatly
improved by the addition of a Gaussian emission line; in this case the
fit had $\chi^2_{\mathrm{dof}} = 0.56$ for 8 degrees of freedom (see
Figure~\ref{fig:spectra}). The probability of obtaining such a value
of $\chi^2$ or larger under the hypothesis that the model is correct
is $P(\chi^2 \ge 4.75 ) = .784$. The line energy was
$E=13.09\pm0.25$~keV. Note that a line at a similar energy was found
by \citet{gkw02} in the first burst discovered from this source.

To firmly establish the significance of this feature, we performed the
following Monte Carlo simulation in \texttt{XPSEC}. We generated 10000
fake spectra drawn from a simple blackbody model having the same
background and exposure as our data set. We fit the simulated data to
a blackbody model and to a blackbody plus emission line model and
compared the $\chi^2$ difference between the two.  To ensure we were
sensitive to narrow lines when fitting our blackbody plus emission
line model we stepped through different line energies from 2 to 30~keV
in steps of 0.2 keV and refit our spectrum holding the line energy
fixed and recorded the lowest $\chi^2$ value returned. In our
simulations only 11 events had a $\chi^2$ difference greater or equal
to the one found from our data.  Thus, the probability of obtaining a
spectral feature of equal significance by random chance is $\sim
0.0011$. The significance of the spectral feature reported for this
source by \citet{gkw02} at this energy was $\sim 0.0008$. Since these
were independent measurements, the probability of finding two spectral
features at the same energy by random chance is $\sim 8.8\times
10^{-7}$, thus the emission line at $\sim$13~keV is genuine.

\subsubsection{Burst Energetics}
\label{sec:flux and fluence}

In order to compare the energetics of this burst to those emitted in
2001 we measured its peak flux and fluence. The first step in this
analysis was to model the background count rate. First we extracted an
instrumental background for the entire observation using
\texttt{pcabackest}. The function \texttt{pcabackest} can only estimate the
background rate every 16~s seconds, so we interpolated to finer
resolution by modeling the background rate as a 5th order
polynomial. We then added the average non-burst count rate to this
model. We estimated this value by subtracting our interpolated
\texttt{pcabackest} model from our data and then measuring the average
count rate over the same interval used to estimate the background in
the spectral evolution analysis. The 2--20~keV peak flux was
determined from the event data using a box-car integrator of width
$1/\Delta t$. We used $\Delta t =64$~ms and $\Delta t = t_r$
\citep[for details on the flux calculation algorithm,
see][]{gkw04}. At each step of the boxcar we subtracted the total
number of background counts as determined by integrating our
background model over the boxcar limits. To convert our flux
measurements from count rates to CGS units we extracted spectra whose
limits were defined by the start and stop time of the boxcars.  For
each flux measurement we extracted a spectrum for the region of
interest, a background spectrum and a response matrix, similar to what
was done for the spectral evolution analysis. Each spectrum was fit to
a photolectrically absorbed blackbody using \texttt{XSPEC} in order to
measure the 2--20~keV flux in CGS units. The 2--20~keV total fluence
was determined by integrating our background-subtracted time
series. If the burst had emitted all of its energy during the
observation, the integrated burst profile would eventually plateau.
However, this is not what we observed. The integrated burst profile
was still steadily rising even at the end of our observation,
indicating that our observation finished before catching the end of
the burst.  Thus, we can only set an upper-limit on the total
2--20~keV fluence; see Table~\ref{ta:burst}.  To convert our fluence
upper limit from counts to CGS units the exact procedure was followed
as for the peak flux measurements.

\subsubsection{Pulsed Flux Measurements}

\begin{figure}
\plotone{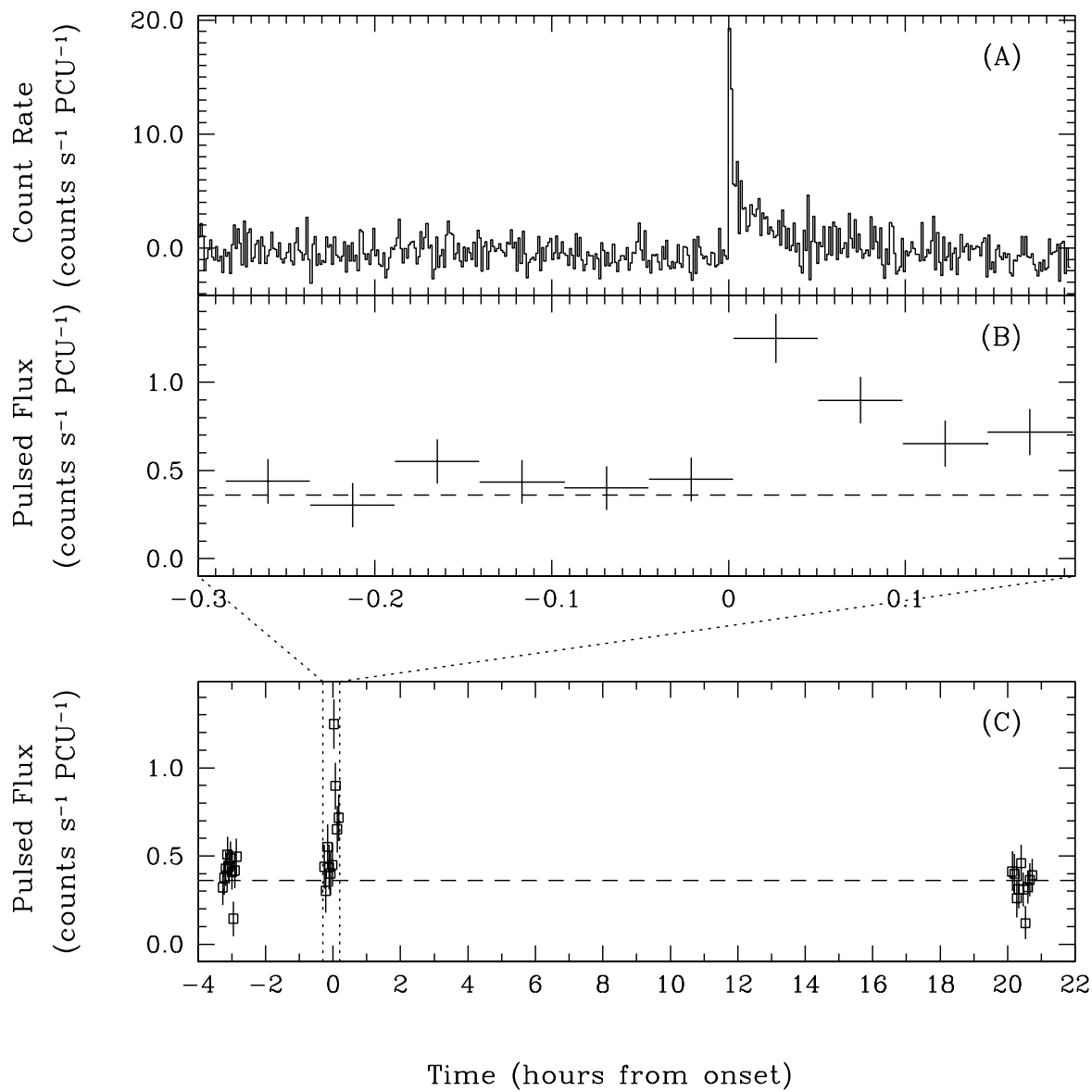} \figcaption{A: Burst time series in the 2--20~keV
band, binned with 4-s time resolution. B: Pulsed flux in the 2--10~keV band
during the observation. The dashed line represents the average
\textit{quiescent} pulsed flux as measured from neighboring observations. C:
Same as above except for a longer baseline. \label{fig:flux}}
\end{figure}

Magnetar candidates have been observed to be highly flux variable,
which is why we regularly monitor the pulsed flux of this source
\citep[see][for a detailed discussion of pulsed flux calculations for
\tfe]{gk04}.  The pulsed flux during the entire observation in which
the burst occured was not significantly higher than in neighboring
observations.  However, in some AXPs and SGRs, short time-scale ($\ll
1000$~s) abrupt changes in pulsed flux have been observed in
conjunction with bursts
\citep[e.g.][]{lwg+03,wkt+04,wkg+05}. Motivated by such observations
we decided to search for short-term pulsed flux enhancements around
the time of the burst from \tfe. We broke the observation into 10
intervals and calculated the pulsed flux for each. In order to avoid
having the burst spike biasing our pulsed flux measurements we removed
a 4~s interval centered on the burst peak. A factor of 3.5 increase in
pulsed flux can be seen in the tail of the burst (see
Fig.~\ref{fig:flux}). This coupling between bursting activity and
pulsed flux establishes that \tfe\ is definitely the burst source.

\subsection{Imaging X-ray Observations}

Following the discovery of a new burst from the direction of \tfe, we
triggered observations of the source with imaging X-ray
telescopes. The AXP was observed once with \xmm\ on 2004 July 8 for 33
ks and twice with \cxo\ on 2004 July 10 and 15 for 29 and 28 ks,
respectively.  Simultaneous \xte\ observations were performed during
the \xmm\ and the first \cxo\ observation to assist in the
identification of bursts. For scheduling reasons the second \cxo\
observation could not be coordinated with simultaneous \xte\
observations.

The \xmm\ data were processed using the {\it XMM-Newton} Science
Analysis
System\footnote{http://xmm.vilspa.esa.es/external/xmm\_sw\_cal/sas.shtml}
(SAS) v6.0.0.  The scripts {\tt epchain} and {\tt emchain} were run on
the Observation Data Files for the PN and MOS data, respectively.  For
the \cxo\ data, we started from the filtered event 2 list for all
results presented here.  Standard analysis threads were followed to
extract filtered event lists, light curves and spectra from the
processed data. See below for more details.  The approximate count
rates for the \xmm\ PN observation was $2.69\pm 0.01$~counts~s$^{-1}$
in the 0.5--12~keV band. The first and second \cxo\ observation had a
count rate of $1.344\pm 0.007$~counts~s$^{-1}$ and $1.302\pm
0.007$~counts~s$^{-1}$, respectively, in the 0.5--10~keV band.

\subsubsection{Burst Search}

The \xmm\ PN camera and \cxo\ ACIS detectors (S3 chip) were operated
in similar modes ({\tt TIMING} for PN and {\tt CC mode} for ACIS) to
optimize the time resolution in order to search for short X-ray bursts
(5.96 ms for PN and 2.85 ms for ACIS).  Filtered source event lists
(0.5$-$12.0 keV for PN and 0.5$-$10.0 keV for ACIS) were extracted for
each observation to create light curves at three different time
resolutions: 1, 10 and 100 times the nominal time resolution of the
data.  No bursts were found in any of the light curves.  Moreover, no
bursts were seen within the \xte\ PCA data of the simultaneous
observations of \tfe.  Aside from the regular X-ray pulsations, the
intensity of \tfe\ does not vary significantly within these
observations.

\subsubsection{Persistent Emission Properties}
\label{sec:pep}

Motivated by the sometimes dramatic changes in the persistent, pulsed
X-ray emission of SGRs and AXPs following burst activity
\citep[e.g.][]{wkt+04}, we investigated both the spectral and temporal
properties of the X-ray emission from \tfe\ using the \xmm\ and \cxo\
data.  Admittedly, only a single, relatively weak burst was observed
from \tfe\ on 2004 June 29, so significant changes in these properties
are not necessarily expected.  However, this particular source has
shown significant variability in its pulsed flux and pulsed fraction
over the last several years apparently independent of strong burst
activity \citep{gk04,mts+04}, so searching for continued evolution in
these properties is of interest.

The {\tt TIMING} mode for PN data is not yet well calibrated for
spectral analysis, so we accumulated a spectrum from the MOS1 camera
which was operated in {\tt SMALL WINDOW} mode.  The source spectrum
was extracted from a circular region centered on the AXP with a radius
of 35$^{\prime \prime}$.  A background spectrum was extracted from a
circular (radius = 80$^{\prime \prime}$), source free region on
CCD~\#3, closest to the center of the field of view.  The calibration
files used to generate the effective area file and response matrix
were downloaded on 2004 August 3.  The source spectrum was grouped to
contain no fewer than 25 counts per channel and fit using
\texttt{XSPEC}. We obtained a satisfactory fit to the energy spectrum
(0.3$-$12.0 keV) using the standard blackbody plus power-law (BB+PL)
model.  Fit parameters are given in Table~\ref{ta:spectra}.

For the \cxo\ observations, source spectra were extracted from a
rectangular region centered on the source with a dimension along the
ACIS-S3 readout direction of 16 pixels ($\sim$8$^{\prime
\prime}$). Background spectra were extracted from 40 pixel wide
rectangular regions on either side of the source with a gap of 7
pixels in between the source and background regions.  Effective area
files and response matrices were generated using CALDB v2.23.  Similar
to the \xmm\ spectral analysis, the \cxo\ spectra were grouped and fit
to the BB+PL model.  We were unable to obtain a satisfactory fit to
either \cxo\ data set due to the presence of an emission line at 1.79
keV.  This feature is instrumental in origin caused by an excess of
Silicon fluorescence photons recorded in the CCD \citep{mskk03}.  To
avoid this feature and calibration uncertainties between 0.3 and 0.5
keV that give rise to large residuals in this range, we restricted our
spectral fits to 0.5$-$1.67 and 1.91$-$10.0 keV. See
Table~\ref{ta:spectra} for fit parameters.

Using the PN data from \xmm\ and the ACIS data from \cxo, we measured
the root-mean-square (rms) pulsed fraction of \tfe\ during each of the
three observations for different energy bands.  For the PN data, a
source event list was extracted from a 5.6 pixel wide rectangular
region centered on the
AXP\footnote{http://wave.xray.mpe.mpg.de/xmm/cookbook/EPIC\_PN/timing/timing\_mode.html}
and the times were corrected to the solar system barycenter.  A
background event list was extracted from two 10 pixel wide rectangles
on either side of the source with a 10 pixel gap between the source
and background regions.  The same source and background regions used
for the spectral analysis of the \cxo\ data were used here for the
pulse timing analysis.  The \cxo\ photon arrival times were corrected
for instrumental time
offsets\footnote{http://wwwastro.msfc.nasa.gov/xray/ACIS/cctime/} and
to the solar system barycenter.  Best-fit frequencies were measured
independently for each observation and found to be consistent with the
more precise \xte\ spin ephemeris.  We constructed
background-subtracted folded pulse profiles for each observation for
different energy ranges and measured the rms pulsed fraction following
\citet{wkt+04} using the Fourier power from the first 3 harmonics.
The pulsed fraction increases with energy from 46.6$\pm$0.5\% in the
0.5$-$1.7~keV range to 56.8$\pm$1.0\% in the 3.0$-$7.0~keV range.
The 2.0$-$10.0 keV pulsed fractions are listed in
Table~\ref{ta:spectra}.

Comparing our results to those of \citet{mts+04}, we find that both
the flux and the pulsed fraction are intermediate between the \xmm\
observations in 2000 December and 2003 June (Figure~\ref{fig:pulsed
flux}).  The definition of pulsed fraction introduced by
\citet{mts+04} is significantly different than ours. Therefore, we
analyzed each of the archival \xmm\ data sets for this source in the
same manner as described above for the 2004 data set.  Similar to the
behavior seen in 2004, we find that the pulsed fraction increases
significantly with energy within each archival data set.  The average
pulsed fractions during these observations, however, are significantly
different than in 2004.  We measure rms pulsed fractions (2.0$-$10.0
keV) of 76.0 $\pm$ 2.3\% and 43.9 $\pm$ 0.4\% for the respective
observations.  The observation in 2000 took place well before the
onset of the first pulsed flux flare \citep{gk04} and the two bursts
seen near the peak of that flare \citep{gkw02}.  The observation in
2003 took place during the decay of the second flare which peaked one
year earlier.  We conclude that both the flux and pulsed fraction that
changed so drastically during these extended flares appear to be
returning to their ``nominal'' pre-flare levels.  It is interesting to
note that the pulsed fraction decreases during these flares showin a
possible anti-correlation with the phase-averaged flux. Thus, the
relative increase in the phase-averaged flux is actually much larger
than the increase in pulsed flux seen in the \xte/PCA light curve
(i.e.\ the phase-averaged flux time history may have exhibited a
stronger peak).

\section{Discussion}
\label{sec:discussion}

We have discovered the longest, most luminous and most energetic burst
from \tfe\ thus far.  The short-term pulsed flux enhancement at the
time of the burst establishes that \tfe\ is definitely the burst
source and in all likelihood was the source of the 2001 bursts as
well.

An interesting property of all three bursts from \tfe\ is that they
occur preferentially at pulse maximum.  A similar trend was found for
the transient AXP \ett\, for which four bursts occurred near pulse
maximum \citep{wkg+05}.  Furthermore, in a major outburst from AXP
\tfn\ involving $\sim$80 bursts, \citet{gkw04} found that bursts
occurred preferentially at pulse phases for which the pulsed emission
was high \citep[note that the pulse profile of \tfn\ is double-peaked
as opposed to the quasi-sinusoidal profiles of \tfe\ and \ett,
see][]{gk02}.  SGR bursts on the other hand show no correlation with
pulse phase.  \citet{pal02} found that hundreds of bursts from
SGR~1900$+$14 were distributed uniformly in phase.  However as
discussed by \citet{gkw04}, \citet{wkg+05} and below, this is not the
only difference between SGR and AXP bursts.

If AXP bursts do occur at specific pulse phases then they must be
associated with particularly active regions of the star.  This would
imply that AXPs burst much more frequently than is observed, but the
bursts go unseen because they are beamed away from us.  However, even
if a burst is missed, it may still leave two characteristic
signatures.  One is a very long tail: those observed in \tfe\ and
\ett\ lasted several pulse cycles. Second, short-term increases in
pulsed flux like those observed in this paper would be an indication
of a burst whose onset went unobserved. A search for ``naked tails''
or short time scale pulsed flux enhancements could in principle
demonstrate the existence of missed bursts.

The very long tail ($>699$~s) of the burst reported here makes it very
similar to one burst observed from \tfe, some of the bursts seen in
AXP \tfn, and to those observed in \ett.  Their long durations set
these bursts apart from the brief $\sim0.1$~s burst observed in
SGRs. As argued by \citet{wkg+05}, bursts from \tfe\, \ett\ and \tfn,
which have very long tails and occur close to pulse maximum, might
constitute a new class of bursts unique to AXPs.  Although there were
two bursts with extended tails observed from SGR~1900$+$14, many of
their properties differ from those of the long-duration AXP
bursts. The first extended-tail SGR burst occurred on 1998 April 29
\citep{isw+01} and the second on 2001 April 28 \citep{lwg+03}. In both
cases an obvious distinction could be made between the initial
``spike'' and the extended ``tail'' component of the burst. For the
AXP bursts which have a fast rise and smooth exponential decay
morphology, there is no point which clearly marks the transition
between initial spike and extend tail. Furthermore in both of these
extended-tail SGR bursts, the majority of the energy was in the
initial spike, not the tail.  In fact, from the time of their peaks to
$\sim 1\%$ of their total duration, $\sim 98\%$ of their total energy
was released. By contrast, in the AXP bursts, virtually all the energy
is in what would be considered the tail component. For the burst
reported here, from the time of its peak to $\sim 1\%$ of its total
duration $< 37\%$ of its total energy was released. Also, unlike the
long-duration AXP bursts which all occurred near pulse maximum, the
extended-tail SGR bursts occurred 180\degr\ apart in pulse phase,
i.e., the first burst occurred near pulse maximum and the second burst
near pulse minimum \citep{lwg+03}. Last, the long-tailed SGR bursts
only occurred following high-luminosity flares: the first followed the
1998 August 27 SGR~1900$+$14 event and the second followed the 2001
April 18 event \citep{gmf+01}. No such high-luminosity flares have
ever been observed in an AXP.

From its earliest stages, the magnetar model put forward by
\citet{td95} offered a viable burst mechanism for SGRs and AXPs.  In a
magnetar, the magnetic field is strong enough to crack the crust of
the neutron star.  The fracturing of the crust disturbs the magnetic
field foot-points and releases an electron-position-photon fireball
into the magnetosphere.  The fireball is trapped and suspended above
the fracture site by closed field lines. The suspended fireball heats
the surface, and in the initial version of this model, it was
suggested that the burst duration would be comparable to the cooling
time.  In more recent work it has been suggested that burst durations
can be extended by orders of magnitude via vertical expansion of the
surface layers \citep{tlk02} or deep crustal heating \citep{let02}.
The surface fracture mechanism can explain the very long durations of
the bursts observed from \tfe\ and \ett.  Furthermore this mechanism
also provides an explanation for the phase dependence of the AXP
bursts, since the fracture sites are thought to be preferentially
located near the magnetic poles.  Hence the bursts would be associated
with a particular active region on the surface, resulting in a
correlation with pulse phase.

\citet{lyu02} proposed another burst emission mechanism within the
framework of the magnetar model. He suggested that the bursting
activity of AXPs and SGRs is due to the release of magnetic energy
stored in non-potential magnetic fields by reconnection-type events in
the magnetosphere.  In this model, bursts occur at random phases
because the emission site is high in the magnetosphere. Hence we
observe all bursts.  This mechanism will produce harder and shorter
bursts as compared to the ones due to surface fracturing.  Softer and
longer bursts are achieved by a combination of reconnection and a
small contribution from surface cooling, as energetic reconnection
events will precipitate particles which will heat the surface.  Since
there is a duration-fluence correlation in both SGR \citep{gkw+01} and
AXP \citep{gkw04} bursts, this model suggests that the shorter
(less-luminous) bursts are harder than the longer (more luminous)
bursts.  A hardness-fluence correlation was found for SGR bursts
\citep{gkw+01}, but an anti-correlation was found for the 80 bursts
from AXP \tfn\ \citep{gkw04}.  It should also be noted that for \tfe\
and \ett\ the more energetic bursts are the hardest, although only
three bursts have been observed thus far.  Hence the aspects that
differentiate the surface-cooling model from the reconnection model
for bursts seem to be the same aspects that separate the canonical SGR
bursts from the long-duration AXP bursts. In the surface-cooling model
one expects longer durations, a correlation with pulse phase and a
fluence-hardness anti-correlation.

It is possible that both mechanisms (surface and magnetospheric) are
responsible for creating AXP and SGR bursts, but that magnetospheric
bursts are more common in SGRs. This is not unreasonable if we
consider the twisted-magnetosphere model proposed by \citet{tlk02}. In
this extension to the magnetar model, \citet{tlk02} suggested that the
highly twisted internal magnetic field of a magnetar imposes stresses
on the crust which in turn twist the external dipole field. The
twisted external fields induce large-scale currents in the
magnetosphere. The inferred dipole magnetic field strengths,
luminosity and spectra of SGRs all suggest that the global ``twists''
of their magnetic fields are greater than those of the AXPs. If the
external fields of SGRs are much more ``twisted'' than those of the
AXPs, that would make them more susceptible to reconnection type
events in their magnetosphere. Furthermore, if SGRs have stronger
magnetic fields than the AXPs, then they would be less susceptible to
surface fracture events because at high field strengths the crustal
motions are expected to become plastic.

What could make a burst tail last longer in AXPs? One possibility is
that the energy is released very deep in the crust and the energy
conducts to the surface on very long time scales.  In this scenario
most of the heat is absorbed by the core, which could explain why AXP
bursts are in general dimmer than SGR bursts. In order for this model
to explain both the extended-tail SGR bursts and the long-duration AXP
bursts one can imagine a hybrid scenario in which a sudden twist
occurs, which deposits energy both in the magnetosphere and deep in
the crust. The energy deposited in the magnetosphere gives rise to a
spike and the energy deposited deep in the crust conducts to the
surface on longer time scales. The reason why no spike is seen in the
AXPs is because the magnetospheric component is small. This reasoning
applies to AXPs \tfe\ and \ett\ but it is not surprising that AXP
\tfn\ exhibits both types of bursts, because as argued by
\citet{wkt+04} the best explanation for the bursts and contemporaneous
flux, pulse profile and spin-down variability from this source was
through a catastrophic event that simultaneously impacted the interior
and the magnetosphere of the star.

The spectrum of the burst reported here is intriguing.  First,
although this burst is much harder given its luminosity when compared
to SGR bursts (the SGRs show a luminosity-hardness anti-correlation),
its spectral softening is very similar to those of the extended-tail
bursts of SGR~1900$+$14 \citep{isw+01, lwg+03}.  Second, the evidence
of a spectral feature at $\sim13$~keV makes this burst very similar to
the first burst detected from \tfe\ \citep{gkw02}.  The probability of
observing an emission line at $\sim$13~keV in both bursts is
exceedingly small, thus the line is almost certainly intrinsic to the
source and has important implications. If it is a proton-cyclotron
line then it allows us to calculate the surface magnetic field
strength, $B = {m_p c E}/{\hbar e} = 2.1\times
10^{15}\left({E}/{\mathrm{13\ keV}}\right)~\mathrm{G}$, where $m_p$
and $e$ are the proton mass and charge, respectively, and $E$ is the
energy of the feature.  This value for the magnetic field strength is
greater than that measured from the spin-down of the source, but the
spin-down is only sensitive to the dipolar component of the field and
it is plausible that the field is multipolar. It should be noted that
this surface magnetic field strength estimation assumes the burst was
a surface cracking phenomenon.  An open question is why bursts from
certain magnetar candidates exhibit spectral features and others do
not.  \tfn\ exhibited over 80 bursts and no spectral features were
seen, whereas \tfe\ showed only three bursts with 2 spectral features
seen. \citet{wkg+05} reported a spectral feature at approximately the
same energy in a burst from AXP \ett\ at a higher significance level
than both of the \tfe\ spectral features.

We have argued that the bursts from \tfe\ are very similar to those of
\ett.  Interestingly, the two sources show other similarities.  They
both have sinusoidal profiles while all other AXPs exhibit rich
harmonic content in their pulse profiles \citep[see][]{gk02,
ggk+02}. They have shown long-lived ($>$ months) pulsed flux
variations which are not due to cooling of the crust after the
impulsive injection of heat from bursts.  Furthermore, their timing
properties are the most reminiscent of the SGRs.

The pulsed fraction decrease we have observed lends further evidence
that the pulsed flux variations observed by \citet{gk04} represent a
new phenomenon seen exclusively in the AXPs.  The fact that the pulsed
fraction decreased as the pulsed flux increased without any pulse
morphology changes implies that there was a greater fractional
increase in the unpulsed flux than in the pulsed flux, in agreement
with what was found by \citet{mts+04}.  Such a flux enhancement cannot
be attributed to a particular active region.  Thus we can rule out the
flux enhancements were due to the injection of heat from bursts that
were beamed away from us, because in that scenario one would expect a
larger fractional change in pulsed flux than in total flux.  Indeed,
during the burst afterglow pulsed flux enhancement in SGR~1900$+$14,
\citet{lwg+03} found a pulsed flux and pulse fraction increase.

\section{Conclusions}
\label{sec:conclusions}

We reported on the discovery of the latest burst from the direction of
AXP \tfe. This burst was discovered as part of our long-term
monitoring campaign of AXPs with \xte. Contemporaneously with the
burst we discovered a pulsed-flux enhancement which unambiguously
identified \tfe\ as the burst's origin.  The clear identification of
\tfe\ as the burster in this case argues that it was indeed the emitter
of the two bursts discovered from the direction of this source  in 2001,
as already inferred by \citet{gkw02}.

All three bursts from \tfe\ can only be explained within the context
of the magnetar model, however many of their properties differentiate
them from canonical SGR bursts. This and the first burst discovered
from this source had very long-tails, $>699$~s and $\sim$51~s
respectively, as opposed to the $\sim$0.1~s duration SGR bursts. Two
ks-long SGR bursts have been reported but we argued that they were a
very different phenomenon \citep{isw+01, lwg+03}. Specifically the
extended-tail SGR bursts had very energetic initial spikes and the
long tails were argued to be the afterglow of this initial injection
of energy. However, in the AXP bursts no such spikes are present; in
fact, most of the energy is in what would be considered the tail.  All
three bursts from \tfe\ occurred near pulse maximum, as opposed to the
SGR bursts which are uniformly distributed in pulse phase. All of the
bursts discovered from AXP \ett\ and a handful of the bursts from AXP
\tfn\ share many properties with the \tfe\ bursts and, as argued by
\citet{wkg+05}, long-tailed bursts (with no energetic spikes) which
occur near pulse maxima might comprise an new burst class thus far
unique to AXPs.

The spectral evolution of this burst was very similar to the
extended-tail SGR bursts with the trend going from hard to
soft. However, at one part of the bursts tail there was an unusual
spectral feature at $\sim$13~keV; a similar feature was discovered in
the first burst from \tfe\ and in a very high signal-to-noise burst
from \ett\ \citep{gkw02,wkg+05}. If features such as these are
confirmed they may provide direct estimates of the neutron star's
magnetic field especially if harmonic features can be positively
identified.

The differences between the long- and short-duration bursts might be
due to separate emission mechanisms. Two burst mechanisms have been
proposed within the magnetar model: surface fracture and
magnetospheric reconnection \citep{td95,lyu02}. We argued that this
burst is more likely a surface fracture event which agrees with the
conclusion reached by \citet{wkg+05} for the long-tailed bursts from
\ett. If in fact the two classes of bursts are due to emission
mechanisms operating in two distinct regions (near the surface and in
the upper magnetosphere), then AXP bursts provide  opportunities to
probe the physics of these separate regions of a
magnetar.

\acknowledgments 
We are very grateful to Maxim Lyutikov for valuable
comments, suggestions and useful discussion.  This work is supported
from NSERC Discovery Grant 228738-03, NSERC Steacie Supplement
268264-03, a Canada Foundation for Innovation New Opportunities Grant,
FQRNT Team and Centre Grants, the Canadian Institute for advanced
Research, SAO grant GO4-5162X and NASA grant NNG05GA54G.  V.M.K. is a
Canada Research Chair and Steacie Fellow.  This research has made use
of data obtained through the High Energy Astrophysics Science Archive
Research Center Online Service, provided by the NASA/Goddard Space
Flight Center.

\begin{deluxetable}{lc}
\tablecolumns{2}
\tablecaption{Burst Timing and Spectral Properties \label{ta:burst}}
\tablehead{\multicolumn{2}{c}{Temporal Properties}  }
\startdata
Burst day, (MJD) & 53185\\
Burst start time, (UT) &  6:52:33.63(18)  \\
Burst rise time, $t_r$ (ms) & 18.2$^{+5.8}_{-4.4}$  \\
Burst duration, $\tnin$ (s) & $>699$ \\
Burst phase &  $-0.078\pm0.016$  \\
\cutinhead{Fluxes and Fluences} 
$\tnin$ fluence\tablenotemark{a} (counts) &  $>5387$\\
$\tnin$ fluence\tablenotemark{a} ($\times 10^{-10}~\mathrm{erg~cm}^{-2}$) & $>330$ \\
Peak flux for 64~ms\tablenotemark{a} ($\times 10^{-10}~\mathrm{erg~s^{-1}~cm}^{-2}$)    & $59\pm9$   \\
Peak flux for $t_r$~ms\tablenotemark{a} ($\times 10^{-10}~\mathrm{erg~s^{-1}~cm}^{-2}$) &  $105\pm20$ \\
\cutinhead{Spectral Properties} 
\underline{Power law:} & \\
Power law index & $1.06^{+0.14}_{-0.12}$  \\
Power law flux ($\times 10^{-11}~\mathrm{erg~s^{-1}~cm^{-2}}$)  & $2.6\pm0.9$  \\
Reduced $\chi^2$/degrees of freedom & 1.00/62  \\
 & \\
\underline{Blackbody:} &  \\
$kT$ (keV) & $2.99^{+0.25}_{-0.23}$  \\
Blackbody flux ($\times 10^{-11}~\mathrm{erg~s^{-1}~cm^{-2}}$) & $2.4\pm0.3$ \\
Reduced $\chi^2$/degrees of freedom &  0.78/62\\
\enddata
\tablecaption{\label{ta:burst}}
\tablenotetext{a}{Fluxes and fluences calculated in the 2--20~keV band.}
\end{deluxetable}

\begin{deluxetable}{cccc}
\tablewidth{\columnwidth}
\tablecolumns{4}
\tablecaption{Phase-averaged spectral fit parameters and pulsed fractions of 
\tfe.\label{ta:spectra}} 
\tablecolumns{4}
\tablehead{\colhead{Parameter\tablenotemark{a}}  	      &  \colhead{July 8}   &  \colhead{July 10}  & \colhead{July 15}}   
\startdata
$N_{H}$ (10$^{22}$ cm$^{-2}$) & 1.18(4)   & 1.18(4)   & 1.23(5)   \\
$kT$ (keV)		      & 0.619(14) & 0.585(12) & 0.585(12) \\
$\Gamma$   		      & 3.30(9)   & 3.08(11)  & 3.23(12)  \\
Flux\tablenotemark{b} (10$^{-12}$ ergs cm$^{-2}$ s$^{-1}$)       & 7.51 & 7.91 & 7.51 \\
Unabs Flux\tablenotemark{c} (10$^{-12}$ ergs cm$^{-2}$ s$^{-1}$) & 9.07 & 9.56 & 9.20 \\
$\chi^2$/dof                  & 307/306   &  293/286  & 306/278   \\
Pulsed Fraction\tablenotemark{d}         & 0.561(7)   & 0.563(9) & 0.551(10) \\
\enddata

\tablenotetext{a}{Numbers in parentheses indicate the 1$\sigma$
uncertainty in the least significant digits of the spectral
parameter. Note that these uncertainties reflect the 1$\sigma$ error
for a reduced $\chi^2$ of unity.}  \tablenotetext{b}{Observed flux
from both spectral components 2$-$10 keV.}
\tablenotetext{c}{Unabsorbed flux from both spectral components 2$-$10
keV.}  \tablenotetext{d}{Rms pulsed fraction (2.0$-$10.0 keV)
following the definition in Woods et al.  (2004).}
\end{deluxetable}

\begin{figure}
\plotone{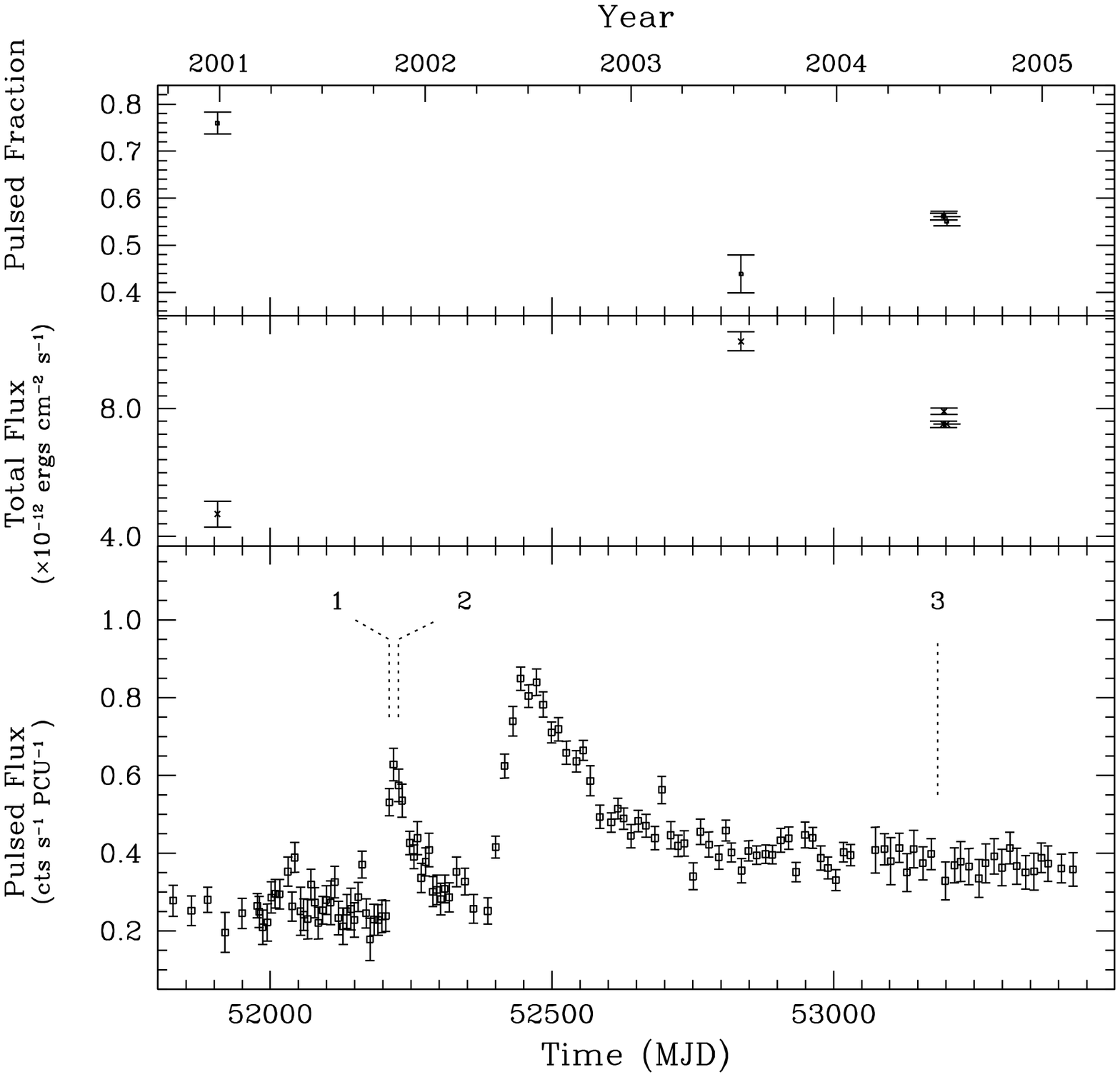} \figcaption{\tfe's total and pulsed flux
evolution. Top: 2--10~keV pulsed fraction as measured by \xmm\ and \cxo\ 
\citep[see][for our particular definition of pulsed fraction]{wkt+04}.
Middle: 2--10~keV total flux as measured by \xmm\ and \cxo. Bottom:
2--10~keV pulsed flux as measured by \xte\ \citep[for details on the analysis
see][]{gk04}. The epochs of the three bursts observed from \tfe\ are
indicated by their respective numbers. \label{fig:pulsed flux}}
\end{figure}

\end{document}